\documentclass[letterpaper,journal]{IEEEtran}

\usepackage{amsmath,amsfonts,amssymb}
\usepackage{algorithmic}
\usepackage{algorithm}
\usepackage{array}
\usepackage[caption=false,font=normalsize,labelfont=sf,textfont=sf]{subfig}
\usepackage{textcomp}
\usepackage{stfloats}
\usepackage{url}
\usepackage{verbatim}
\usepackage{graphicx}
\usepackage{cite}
\usepackage{xcolor}
\colorlet{blue}{black}

\usepackage{booktabs}
\usepackage{siunitx}
\usepackage{makecell}
\usepackage{placeins}

\begin{document}

\title{Thermal-Mass Coupled Modeling and Optimal Scheduling of Aluminum Smelting Loads for Safe Demand Response Participation}

\author{Yan Shen, \IEEEmembership{Student Member, IEEE},
        Ruike Lyu, \IEEEmembership{Student Member, IEEE},
        Shuoqi Wang,
        Jing Dai,
        Hongye Guo, \IEEEmembership{Member, IEEE},
        and Chongqing Kang, \IEEEmembership{Fellow, IEEE}%
\thanks{Yan Shen, Ruike Lyu, Hongye Guo, and Chongqing Kang are with the State Key Laboratory of Power System Operation and Control, Department of Electrical Engineering, Tsinghua University, Beijing 100084, China (e-mail: hyguo@tsinghua.edu.cn).}
\thanks{Shuoqi Wang is with Shanghai Lumin AI Co., Ltd. (e-mail: shuoqi.wang@lumin-ai.cn).}
\thanks{Jing Dai is with the Energy Internet Research Institute, Tsinghua University, Beijing 100085, China (e-mail: jingdai@tsinghua.edu.cn).}
\thanks{Corresponding authors: Hongye Guo and Jing Dai.}}

\markboth{IEEE Transactions on Smart Grid,~Vol.~XX, No.~X, Month~202X}%
{Shen \MakeLowercase{\textit{et al.}}: Thermal-Mass Coupled Modeling and Scheduling of ASLs}

\maketitle

\begin{abstract}
Aluminum smelting loads (ASLs) offer substantial flexibility potential, yet their actual participation in demand response remains conservative owing to the risk of operational failures. Power adjustments disturb the electrolyte temperature and superheat, which may lead to electrolyte freeze-out or yield loss. Existing ASL demand response methods often rely on oversimplified dynamic models of the electrolytic cell's thermal behavior, such as linear models and the Equivalent Thermal Parameter (ETP) model, which may either violate temperature constraints or underestimate the available flexibility. To address this limitation, this paper proposes a Thermal-Mass Coupled (TMC) dynamic model. The model captures the essential physical mechanisms of electrolytic cells, including alumina feed dynamics, ledge phase-change dynamics, and state-dependent current efficiency. {\color{blue}The resulting nonlinear relations are reformulated into a tractable mixed-integer linear programming (MILP) model using Taylor approximations and McCormick-envelope reformulations.} {\color{blue}In the illustrative PBDR and IBDR case studies, the nonlinear reference-model assessment shows that the TMC model achieves net economic improvements 122.4\% and 325.1\% higher, respectively, than the ETP model. The proposed framework can therefore support deeper demand response participation by aluminum smelters while explicitly assessing adherence to operational safety constraints.}
\end{abstract}


\section{Introduction}
\label{sec1}

\IEEEPARstart{I}{n} the context of the global low-carbon transition, the increasing penetration of variable renewable energy (VRE) necessitates enhanced operational flexibility in modern power systems \cite{mohandes_review_2019}. To address VRE volatility, demand-side flexibility offers a cost-effective alternative to building new generation resources and transmission lines \cite{chen_pathway_2021,mohandes_incentive_2021}{\color{blue}; a unified demand-response (DR) framework can support its deployment across economic and technical objectives \cite{mathieu_dr_definition_2025}.} Energy-intensive industries, which account for a significant portion of global electricity consumption, have demonstrated both the technical capability and economic value of providing grid flexibility through price-based and incentive-based demand response programs \cite{zhang_steel_2017,zhang_ancillary_2018,han_interactive_industrial_2025}. As a typical energy-intensive load, aluminum smelting loads (ASLs) possess gigawatt-level flexibility potential, with concentrated and highly electrified energy consumption that enables power adjustments within seconds \cite{eskin_contribution_2022}.

Despite this potential, aluminum smelters' actual participation in demand response remains limited. The main challenge lies in the strong coupling between electrical power input and the electrolytic cell's thermal balance. Safe operation requires keeping both the electrolyte temperature and the superheat within narrow windows (${\approx}\,950\text{--}970\,\si{\celsius}$ and ${\approx}\,5\text{--}15\,\si{\celsius}$, respectively) \cite{reek_power_2015}. The superheat, defined as the temperature difference between the bulk electrolyte and its liquidus point, governs the stability of the side ledge, a protective frozen cryolite layer that insulates the sidewall \cite{taylor_energy_2014}. {\color{blue}Alongside these thermal states, dissolved alumina concentration is an operationally constrained mass state: depletion can cause excessive anodic polarization and anode effects, whereas excessive or poorly dissolved feed can lead to sludge formation and voltage instability \cite{taylor_heat_balance_1984}.} During demand response, power adjustments disturb {\color{blue}this coupled thermal-mass balance}. If the scheduling strategy cannot accurately predict {\color{blue}the resulting state dynamics}, the electrolyte temperature or the superheat may drift outside the {\color{blue}safety limits}. Such violations can trigger electrolyte freeze-out, sidewall corrosion, or anode effects, resulting in substantial yield loss and economic penalties \cite{chesonis_lengthy_2019}. A single cell freeze-out can cascade into a full potline shutdown with {\color{blue}restart costs of tens of millions of USD} \cite{chesonis_lengthy_2019}, making smelters reluctant to commit their flexibility to deep demand response.

Given the complex thermal dynamics and stringent safety constraints, current industrial practice still relies largely on empirical power adjustments and therefore fails to fully exploit the ASL's demand response potential. Early field tests by Alcoa \cite{Todd2009ProvidingRS} showed that, by controlling on-load tap changers and saturable reactors, ASLs can technically provide a regulation capacity of approximately 2--4\%. However, this participation remains limited to short-term frequency regulation so that total energy input stays within a narrow time window. Statistical analysis based on historical data \cite{molina-garcia_modeling_2011} reported a similarly conservative regulation capacity of approximately $\pm 1.5\%$. This estimate, however, assumes a fixed load current and relies solely on internal resistance adjustments. By contrast, industrial investigations \cite{Eisma2016} indicate that much deeper load current adjustments, ranging from 10\% to 30\%, are physically feasible, yet such flexibility is difficult to realize reliably in practice because large power deviations can readily disrupt the cell thermal balance. Recent industrial studies show that shell heat exchanger technology can expand the feasible modulation window \cite{Depree2016TheB,Depree2024Cell}, but its large-scale adoption remains limited by retrofit cost. Consequently, disruption of thermal balance remains a primary barrier to large-scale ASL demand response despite the theoretically available flexibility \cite{golmohamadi_demand-side_2022}.

Fully exploiting ASL demand response flexibility requires moving beyond empirical power adjustments toward optimal scheduling strategies built on an accurate dynamic model of the electrolytic cell's thermal behavior. Existing ASL demand response research has mainly focused on grid-level benefits, including peak shaving \cite{yue_collaborative_2024}, load frequency control \cite{jiang_mpc-based_2015,liao_control_2018,bao_hierarchical_2020}, incentive-based demand response programs \cite{zeng_bi-level_2020}, and ancillary service markets \cite{schafer_model_2019}. However, these studies commonly rely on oversimplified dynamic models for optimization. Linear models \cite{liao_control_2018,yue_collaborative_2024}, for example, assume that temperature changes are proportional to power changes and therefore neglect complex heat-dissipation mechanisms. Such simplifications may produce schedules that violate temperature constraints. The recently proposed Equivalent Thermal Parameter (ETP) model \cite{huang_novel_2024} represents the cell as a simple first-order thermal system, but it still neglects mass-balance coupling, including alumina feeding and ledge phase change. As a result, it may underestimate the available flexibility for demand response.

At the other extreme, high-fidelity multiphysics models based on Computational Fluid Dynamics (CFD) \cite{Allard2019,Ran2023} or Finite Element Methods (FEM) \cite{dupuis_computation_2016,ran_fully-coupled_2024} can accurately resolve the nonlinear electro-thermal fields within the cell, but they typically require minutes or even hours per time step, making them unsuitable for scheduling optimization. A spatially discretized dynamic model based on fundamental thermal-mass balances \cite{wong_studies_2023} substantially improves simulation efficiency, yet its high spatial granularity and strong nonlinearities still pose challenges for fast optimization. There thus remains a gap between the physical fidelity {\color{blue}needed to support safe operation} and the computational tractability required by demand response optimization.

{\color{blue}To address this modeling gap, this paper develops a Thermal-Mass Coupled (TMC) dynamic model and embeds it into an optimal scheduling framework for aluminum smelting loads. Unlike oversimplified Linear and ETP models, the proposed TMC model explicitly represents the coupled thermal-mass state dynamics and associated safety limits, thereby enabling safety-aware and computationally efficient potline-level scheduling.} Although the framework is applicable more broadly, this paper focuses on two representative demand response mechanisms: price-based demand response (PBDR) and incentive-based demand response (IBDR).

The main contributions of this paper are summarized as follows:
\begin{itemize}
\item {\color{blue}We develop a scheduling-oriented Thermal-Mass Coupled (TMC) dynamic model for aluminum smelting electrolytic cells. The model integrates alumina feeding, ledge phase change, composition-dependent liquidus temperature, and state-dependent current efficiency, and represents the resulting safety-relevant state variables as operational constraints for potline-level demand response scheduling.}
\item {\color{blue}We reformulate the resulting nonlinear TMC-based scheduling problem as a tractable Mixed-Integer Linear Programming (MILP) model through local linearization of the quadratic and empirical state relations and McCormick envelopes for bilinear terms. This formulation retains the coupled state constraints and supports computationally efficient, real-time solution at the potline level.}
\item {\color{blue}We evaluate the proposed model in two representative demand response mechanisms, namely PBDR and IBDR, and compare it with linear and ETP-based models using a nonlinear TMC reference-model assessment. In the tested mechanisms, the added thermal-mass coupling yields higher economic value under this assessment, while nonlinear re-simulation evaluates the resulting schedules against the operational safety constraints.}
\end{itemize}

\section{ASL Optimal Scheduling Framework for Demand Response Participation}
\label{sec:scheduling}

This section formulates the ASL optimal scheduling problem as a discrete-time optimization model, covering the physical decision variable, the economic objective functions under two demand response mechanisms, and the shared physical and operational constraints.

\subsection{Aluminum Smelting Process and Decision Variable}

The Hall-Héroult process is the primary production method for aluminum, which involves the electrolysis of alumina ($\text{Al}_2\text{O}_3$) dissolved in a molten cryolite electrolyte. This process is performed in a series of electrolytic cells, collectively known as a potline. A potline is typically supplied with a large DC current ranging from 300 to 500~kA. The power load of a potline can be adjusted primarily through two mechanisms: (i) adjusting the load current, or (ii) adjusting the Anode-Cathode Distance (ACD). {\color{blue}The load current is the common series-level actuator and is therefore the natural decision variable for potline-level demand response. ACD adjustment, by contrast, is a cell-level process-control action that is mainly used to restore or maintain local heat balance during normal operation; jointly optimizing ACD would require per-cell actuation and a more detailed cell-control layer \cite{taylor_advances_2007,shi_new_2021,wong_studies_2023}. This study therefore focuses on current-only scheduling and takes the load current $I_t$ as the sole decision variable. Through the cell electro-thermal relations, $I_t$ determines power consumption, aluminum production, and the internal thermal-mass state evolution used in the subsequent subsections.}

{\color{blue}The representative-cell formulation also assumes homogeneous cell states and parameters across the scheduled potline. This inter-cell homogeneity assumption keeps the plant-level scheduling problem tractable and is consistent with the fact that all series-connected cells receive the same current. In practice, state dispersion (e.g., temperature, superheat, ledge thickness, and alumina concentration) and parameter dispersion (e.g., heat-transfer coefficients, ledge geometry, and current-efficiency coefficients) can reduce the available load-modulation margin. State dispersion affects the safety margin available at the beginning of a scheduling horizon, whereas parameter dispersion affects the subsequent state evolution under the potline current trajectory. The representative-cell schedule should therefore be implemented with conservative safety margins and online monitoring. The use of cell-level information in rolling scheduling is discussed in Section~\ref{sec:implementation}.}

{\color{blue}Beyond these inter-cell variations, the lumped TMC model does not resolve intra-cell spatial gradients, such as local temperature, composition, anode-current, or ledge variations. Spatially discretized models and cell-level control studies can represent or infer these conditions \cite{wong_studies_2023,wong_discretized_2021,shi_multivariable_2020,shi_new_2021}, but are outside the hourly potline-level scheduling model. The use of cell-level information in rolling scheduling is discussed in Section~\ref{sec:implementation}.}

\subsection{Demand Response Mechanisms and Objective Functions}
{\color{blue}This paper considers two representative demand response mechanisms implemented by the aluminum smelter. In Price-Based DR (PBDR), the smelter adjusts its consumption in response to an exogenous time-varying electricity-price signal. In Incentive-Based DR (IBDR), it responds to a notified load-reduction event and receives performance-based compensation. Both mechanisms use the same electro-thermal model and operational constraints. The corresponding objective functions are formulated as follows.}

{\color{blue}\textbf{1) Price-Based DR (PBDR):} Over a horizon of $T$ intervals, each of duration $\Delta t$, the aluminum smelter schedules its aggregate potline power consumption $P_t$ at time $t$ against an exogenous time-varying electricity-price signal $\pi_t^{\mathrm e}$. In the illustrative case study, this signal is the regulated Time-of-Use (ToU) tariff. The formulation can also accommodate other exogenous price signals, including spot-market prices. The price signal is treated as an input; market clearing and price feedback are not modeled.}

{\color{blue}If the aluminum smelter has entered a medium- or long-term electricity contract before operational scheduling, $P_t^{\mathrm{con}}$ denotes its fixed hourly procurement schedule and $\pi_t^{\mathrm{con}}$ denotes the corresponding contract price. When deviations are settled at $\pi_t^{\mathrm e}$, the total electricity settlement cost $C_{\mathrm{elec}}^{\mathrm{settle}}$ can be written as}
\begin{equation}
{\color{blue}C_{\mathrm{elec}}^{\mathrm{settle}}
=
\sum_{t=1}^{T}
\left[
\pi_t^{\mathrm{con}}P_t^{\mathrm{con}}
+
\pi_t^{\mathrm e}\left(P_t-P_t^{\mathrm{con}}\right)
\right]\Delta t .}
\label{eq:procurement_settlement_cost}
\end{equation}
{\color{blue}Equivalently,}
\begin{equation}
{\color{blue}C_{\mathrm{elec}}^{\mathrm{settle}}
=
\sum_{t=1}^{T}\pi_t^{\mathrm e}P_t\Delta t
+
\sum_{t=1}^{T}
\left(\pi_t^{\mathrm{con}}-\pi_t^{\mathrm e}\right)P_t^{\mathrm{con}}\Delta t .}
\label{eq:procurement_settlement_equiv}
\end{equation}
{\color{blue}The second term is fixed when the operating schedule is optimized. It changes the reported settlement cost but does not affect the optimal power trajectory. Thus, $\text{Cost}_{\text{elec}}$ captures the schedule-dependent electricity cost.}
{\color{blue}The PBDR objective therefore minimizes the schedule-dependent electricity cost together with material costs and yield-loss penalties:}
\begin{equation}
{\color{blue}\min \quad \text{Obj}_{\text{PBDR}} = \text{Cost}_{\text{elec}} + \text{Cost}_{\text{Al}_2\text{O}_3} + \text{Loss}_{\text{Al}}}
\label{eq:obj_pbdr}
\end{equation}
\begin{equation}
{\color{blue}\text{Cost}_{\text{elec}} = \sum_{t=1}^{T} P_t \cdot \pi_t^{\mathrm e} \cdot \Delta t}
\label{eq:cost_elec_main}
\end{equation}
\begin{equation}
\text{Cost}_{\text{Al}_2\text{O}_3} = \sum_{t=1}^{T} \dot{m}_{\text{Al}_2\text{O}_3,t}^{\text{feed}} \cdot \text{price}_{\text{Al}_2\text{O}_3} \cdot \Delta t
\label{eq:cost_al2o3_main}
\end{equation}
\begin{equation}
\text{Loss}_{\text{Al}} = \sum_{t=1}^{T} (\dot{m}_{\text{Al}, N} - \dot{m}_{\text{Al}, t}) \cdot \text{price}_{\text{Al}} \cdot \Delta t
\label{eq:loss_al_main}
\end{equation}
{\color{blue}Here, $\dot{m}_{\text{Al}, N}$ is the rated potline aluminum production rate, $\dot{m}_{\text{Al}, t}$ is the actual potline aluminum production rate governed by Faraday's law in \eqref{eq:production}, and $\dot{m}_{\text{Al}_2\text{O}_3,t}^{\text{feed}}$ is the aggregate potline alumina feed rate. The unit prices $\text{price}_{\text{Al}_2\text{O}_3}$ and $\text{price}_{\text{Al}}$ correspond to alumina and aluminum, respectively. These aggregate quantities follow from the homogeneous-cell assumption introduced in Section~\ref{sec:scheduling}.}

{\color{blue}\textbf{2) Incentive-Based DR (IBDR):} The aluminum smelter responds to a notified load-reduction event and receives an additional performance-based incentive for its credited reduction. For each mandatory response period $t\in\mathcal{T}_{\text{mandatory}}$, $P_t^{\mathrm{base}}$ is the baseline load, $P_{\text{target},t}$ is the notified target power, and $\pi_t^{\mathrm{DR}}$ is the DR incentive price. Its electricity cost remains defined by \eqref{eq:cost_elec_main}; the event adds the DR-revenue term to the operating-cost objective:
\begin{equation}
\min \quad \text{Obj}_{\text{IBDR}} = \text{Cost}_{\text{elec}} + \text{Cost}_{\text{Al}_2\text{O}_3} + \text{Loss}_{\text{Al}} - \text{Revenue}_{\text{DR}}
\label{eq:obj_ibdr}
\end{equation}
Specifically, the credited reduction is multiplied by a performance factor $\mu_t$ determined from the response-performance ratio $\gamma_t$ through a stepped scheme that rewards accurate compliance and penalizes deviations:
\begin{equation}
{\color{blue}\text{Revenue}_{\text{DR}} = \sum_{t \in \mathcal{T}_{\text{mandatory}}} (P_t^{\mathrm{base}} - P_t) \cdot \pi_t^{\mathrm{DR}} \cdot \mu_t \cdot \Delta t}
\label{eq:dr_revenue_main}
\end{equation}
\begin{equation}
{\color{blue}\gamma_t = \frac{P_t^{\mathrm{base}} - P_t}{P_t^{\mathrm{base}} - P_{\text{target},t}}}
\label{eq:ratio}
\end{equation}
\begin{equation}
\mu_t = \begin{cases}
0, & \gamma_t < 0.5 \\
0.3, & 0.5 \leq \gamma_t < 0.8 \\
1, & 0.8 \leq \gamma_t < 1.2 \\
0.3, & 1.2 \leq \gamma_t < 1.5 \\
0, & \gamma_t \geq 1.5
\end{cases}
\end{equation}
{\color{blue}The specific step boundaries and penalty values reflect a typical performance-based settlement rule. In the illustrative case study, $P_t^{\mathrm{base}}$ is set to the rated potline power $P_N$ to isolate the effect of cell-dynamic modeling on dispatch performance. In practical settlement, the baseline load may be determined using historical consumption, market rules, or contractual arrangements \cite{mathieu_quantifying_2011,chao_demand_2011}. A baseline-load sensitivity analysis then evaluates how the baseline assumption affects credited response and DR revenue.}}

\subsection{Shared Physical and Operational Constraints}
{\color{blue}Although the economic objectives differ between PBDR and IBDR, the electro-thermal dynamics governing the electrolytic cell remain identical.} The unified constraints can be logically categorized into two primary aspects: internal physical states and external equipment operations.

\textbf{1) Internal Physical State Constraints:} Let $\mathbf{x}_t = [T_{\text{E},t}, T_{\text{super},t}, L_{\text{ledge},t}, C_{\text{Al}_2\text{O}_3,t}]^T$ denote the multi-physics state vector at time $t$. Operational safety is enforced through bounds on these core states together with a terminal recovery constraint that keeps the final state within a prescribed deadband around the initial state. These constraints are abstracted as
\begin{equation}
\mathbf{x}_{\min} \le \mathbf{x}_t \le \mathbf{x}_{\max}
\end{equation}
\begin{equation}
|\mathbf{x}_{T} - \mathbf{x}_{0}| \le \boldsymbol{\epsilon}
\end{equation}
\begin{equation}
\mathbf{x}_{t+1} = f(\mathbf{x}_t, I_t, \mathbf{w}_t)
\end{equation}
\begin{equation}
P_t = g(\mathbf{x}_t, I_t)
\end{equation}
where $\mathbf{x}_{\min}$ and $\mathbf{x}_{\max}$ are the {\color{blue}safety limits on the states}, $\boldsymbol{\epsilon}$ is the terminal deadband, and $\mathbf{w}_t$ represents purely exogenous disturbances such as ambient-temperature fluctuations.

\textbf{2) External Equipment and Operational Constraints:} The load current is restricted by absolute capacity limits, where $I_{\min}$ and $I_{\max}$ denote the absolute allowable current regulation range dictated by the capacity of the rectifier transformers. Moreover, to prevent excessive mechanical wear, dynamic ramp rate limits and regulation action limits must be enforced on the load current through binary indicators:
\begin{equation}
I_{\min} \le I_t \le I_{\max}
\end{equation}
\begin{equation}
-R_{\text{down}} \cdot \lambda_{\text{down},t} \le I_t - I_{t-1} \le R_{\text{up}} \cdot \lambda_{\text{up},t}
\end{equation}
\begin{equation}
\lambda_{\text{up},t} + \lambda_{\text{down},t} + \lambda_{\text{keep},t} = 1
\end{equation}
\begin{equation}
\sum_{t=1}^{T-1} (\lambda_{\text{up},t} + \lambda_{\text{down},t}) \leq N_{\text{rel}}
\end{equation}
where $R_{\text{up}}$ and $R_{\text{down}}$ denote the upward and downward current ramp-rate limits, respectively. The binary variables $\lambda_{\text{up},t}$, $\lambda_{\text{down},t}$, and $\lambda_{\text{keep},t}$ indicate whether the load current is increased, decreased, or kept unchanged at time $t$. The parameter $N_{\text{rel}}$ denotes the maximum allowable number of current regulation actions over the scheduling horizon.

In conventional power-system models, the explicit forms of the nonlinear functions $f(\cdot)$ and $g(\cdot)$ are often oversimplified, which risks violating critical limits during deep power modulation. To formulate a physically safe scheduling problem, we first derive the nonlinear thermal-mass-coupled mechanisms {\color{blue}represented by} these abstract functions in Section III.

\section{Thermal-Mass Coupled Dynamic Modeling of Electrolytic Cell}
\label{sec:cell_model}

{\color{blue}This section assembles a lumped-parameter thermal-mass coupled (TMC) model that maps the load current to the safety-relevant state variables of the electrolytic cell, thereby providing the explicit nonlinear functions and operational constraints required by the scheduling framework in Section~\ref{sec:scheduling}. The individual constitutive relations are well established in the aluminum-electrolysis literature \cite{taylor_energy_2014,wong_studies_2023,kvande_cell_2000,dewing_loss_1991,solheim_liquidus_2016,Robilliard2016,lavoie_review_2016}; the contribution here is to organize these mechanisms into a compact scheduling-oriented state-space model whose key states can be imposed as demand response safety constraints. Compared with ETP models \cite{huang_novel_2024}, which treat the ledge as static and ignore electrolyte composition, the TMC model retains ledge phase change, alumina mass balance, composition-dependent liquidus temperature, and superheat-dependent current efficiency. Compared with high-fidelity CFD/FEM or spatially discretized dynamic models \cite{Allard2019,dupuis_computation_2016,ran_fully-coupled_2024,wong_studies_2023}, it preserves these scheduling-critical states and safety constraints in a compact form for repeated MILP scheduling from the aluminum smelter's perspective rather than maximizing spatial resolution. Table~\ref{tab:model_scope} compares both the modeling scope and the scheduling compatibility of all four approaches.}

\begin{figure}[htbp]
    \centering
    \includegraphics[width=\columnwidth]{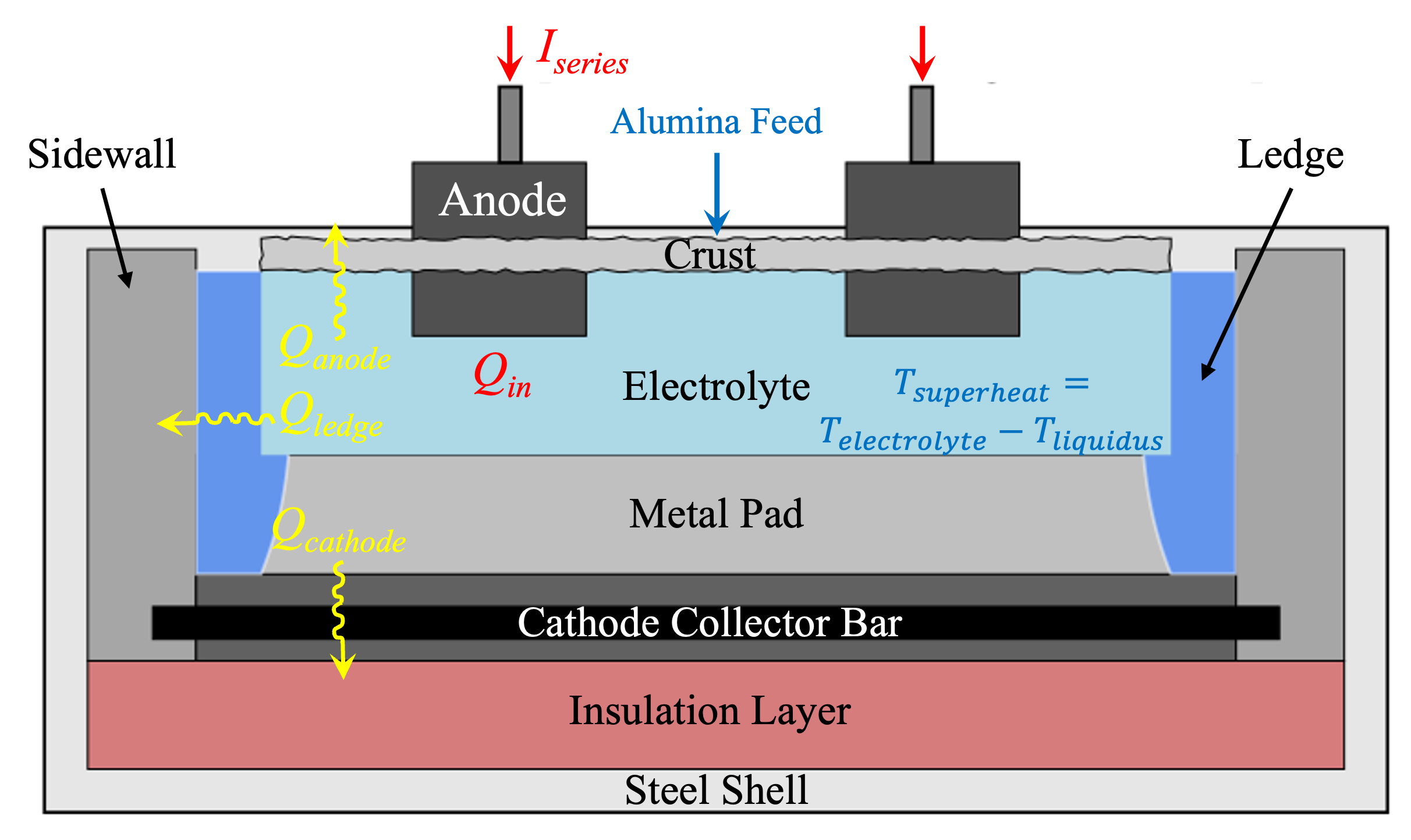}
    \caption{Schematic diagram of the thermal-mass balance in an aluminum electrolytic cell.}
    \label{fig:cell_schematic}
\end{figure}

As illustrated in Fig.~\ref{fig:cell_schematic}, the model comprises three coupled subsystems (power consumption and aluminum production, thermal balance, and mass balance), whose interaction is governed by the superheat. Each thermally distinct region is represented by a single temperature node, and the governing equations are discretized at the scheduling time step. The representative-cell variables retain the same symbols as in Section~\ref{sec:scheduling}; potline-level quantities follow by aggregation under the homogeneous-cell assumption.

\begin{table*}[!t]
\renewcommand{\arraystretch}{1.12}
\caption{{\color{blue}Modeling Scope and Scheduling Compatibility Across Representative Approaches.}}
\label{tab:model_scope}
\centering
\small
\setlength{\tabcolsep}{5pt}
\begin{tabular}{@{}lcccc@{}}
\toprule
Mechanism & \makecell{Linear\\\cite{yue_collaborative_2024}} & \makecell{ETP\\\cite{huang_novel_2024}} & \makecell{Spatially-\\Discretized\\\cite{wong_studies_2023}} & \textbf{TMC} \\
\midrule
Power--current quadratic           & \checkmark & \checkmark & \checkmark & \checkmark \\
Cell energy balance                & algebraic  & first-order & distributed & first-order \\
Ledge as dynamic state             & ---        & ---        & \checkmark & \checkmark \\
State-dependent current efficiency & temperature-linear & --- & \checkmark & superheat-based \\
Composition-dependent liquidus     & ---        & ---        & \checkmark & \checkmark \\
Alumina mass balance               & ---        & ---        & \checkmark & \checkmark \\
\midrule
{\color{blue}Safety-relevant states} & {\color{blue}\makecell{electrolyte\\temperature}} & {\color{blue}\makecell{electrolyte\\temperature}} & {\color{blue}\makecell{spatially discretized\\thermal-mass states}} & {\color{blue}\makecell{temperature; superheat;\\ledge thickness; alumina}} \\
Spatial discretization             & ---        & ---        & \checkmark & --- \\
Sub-hourly dynamics                & ---        & ---        & \checkmark & --- \\
{\color{blue}MILP-compatible form} & {\color{blue}\checkmark} & {\color{blue}\checkmark} & {\color{blue}---} & {\color{blue}\checkmark} \\
\bottomrule
\end{tabular}
\end{table*}

\subsection{Power Consumption and Aluminum Production}
\label{sec:power_production}

For the representative cell, the electrical power consumption is
\begin{equation}
P_t = U_t \cdot I_t
\label{eq:power}
\end{equation}
The cell voltage $U_t$ consists of three main components: (i) the equilibrium potential, (ii) overpotentials, and (iii) ohmic voltage drops \cite{kvande_cell_2000}. The equilibrium potential and overpotentials are independent of current and can be collectively represented as a back electromotive force, $V_{\text{EMF}}$. The ohmic voltage drops arise from resistance in the electrolyte, bubbles, anodes, cathodes, and external superstructures, and are proportional to the current. Since ACD is not considered as a decision variable, the back electromotive force and the total ohmic resistance, $R_{\text{ohm}}$, can be regarded as constant parameters \cite{wong_studies_2023}. Under this assumption, the cell voltage can be expressed as
\begin{equation}
U_t = V_{\text{EMF}} + R_{\text{ohm}} \cdot I_t
\label{eq:voltage}
\end{equation}

The aluminum production rate $\dot{m}_{\text{Al}, t}$ is governed by Faraday's law and is proportional to the load current and the state-dependent current efficiency $\eta_t$:
\begin{equation}
\dot{m}_{\text{Al}, t} = \frac{M_{\text{Al}}}{F \cdot z_{\text{Al}}} \cdot I_t \cdot \eta_t
\label{eq:production}
\end{equation}
where $M_{\text{Al}}$ is the molar mass of aluminum, $F$ is the Faraday constant, and $z_{\text{Al}}$ is the valence of aluminum ions.

To capture the efficiency penalty caused by power-induced thermal deviations, the current efficiency is treated as state dependent rather than constant. Let $T_{\text{super},t} = T_{\text{E},t} - T_{\text{liq},t}$ denote the superheat, where $T_{\text{liq},t}$ is defined in Section~\ref{sec:thermal_balance}. The percentage efficiency loss, $L_t = (1 - \eta_t) \times 100$, follows the empirical correlation of Dewing \cite{dewing_loss_1991}:
\begin{equation}
\log_{10}(L_t) = A_0 + A_1 T_{\text{super}, t} - A_2 C_{\text{AlF}_3,t} - A_3 C_{\text{LiF},t}
\label{eq:current_efficiency}
\end{equation}
where $A_i$ ($i=0,\ldots,3$) are empirical coefficients, and $C_{\text{AlF}_3,t}$ and $C_{\text{LiF},t}$ are electrolyte concentrations in wt.\%, with $\eta_N$ denoting the rated current efficiency under nominal operating conditions.

\subsection{Thermal Balance Dynamics}
\label{sec:thermal_balance}

\textbf{1) Overall Thermal Energy Balance:} The electrolyte and metal pad are lumped into a single control volume at temperature $T_{\text{E},t}$ because of their strong thermal coupling. The metal mass is treated as constant, while the electrolyte mass is coupled to the mass balance. The resulting energy balance is
\begin{equation}
(m_{\text{E},t} c_{\text{E}} + m_{\text{M}} c_{\text{M}}) \frac{d T_{\text{E},t}}{d t} = P_t - Q_{\text{proc},t} - Q_{\text{loss},t}
\label{eq:thermal_balance}
\end{equation}
where $m_{\text{E},t}$ and $m_{\text{M}}$ are the electrolyte and metal masses, $c_{\text{E}}$ and $c_{\text{M}}$ are the corresponding heat capacities, $Q_{\text{proc},t}$ is the process energy, and $Q_{\text{loss},t}$ is the total heat dissipation. The process energy is
\begin{equation}
Q_{\text{proc},t} = (W_{\text{reac}} + W_{\text{heat}}) \dot{m}_{\text{Al},t}
\label{eq:reaction_heating}
\end{equation}
where $W_{\text{reac}}$ and $W_{\text{heat}}$ are the reaction heat and feed preheating energy per unit aluminum production. The total heat dissipation is decomposed as
\begin{equation}
Q_{\text{loss},t} = Q_{\text{anode},t} + Q_{\text{cathode},t} + Q_{\text{ledge},t}
\label{eq:heat_loss_components}
\end{equation}
where $Q_{\text{anode},t}$, $Q_{\text{cathode},t}$, and $Q_{\text{ledge},t}$ denote the anode, cathode, and ledge heat-loss paths, respectively.

\textbf{2) Anode-Path Heat Dissipation:} Heat from the bath is conducted upward through the anode assembly and the alumina covering crust into the gas-filled hood above. Because the thermal mass of the anode carbon is small relative to the control volume, the transient heat storage in the anode is negligible at the scheduling time scale. The heat flux is therefore described by a single lumped thermal resistance:
\begin{equation}
Q_{\text{anode},t} = \frac{T_{\text{E},t} - T_{\text{gas}}}{R_{\text{E-gas}}}
\end{equation}
where $T_{\text{gas}}$ is the gas temperature above the anode and $R_{\text{E-gas}}$ is the bath-to-gas thermal resistance.

\textbf{3) Cathode-Path Heat Dissipation:} The bottom heat loss path runs from the metal pad through the cathode carbon block and underlying insulation to the ambient environment. This path has considerable thermal inertia because the cathode carbon block stores heat while the insulation impedes its release to the environment. Accordingly, the cathode structure is modeled as a first-order RC thermal network, with the carbon block and insulation lumped into a single lining node. Heat flows from the control volume to the lining and then from the lining to the ambient:
\begin{align*}
Q_{\text{cathode},t} &= \frac{T_{\text{E},t} - T_{\text{lining},t}}{R_{\text{M-lining}}} \\
Q_{\text{lining-amb},t} &= \frac{T_{\text{lining},t} - T_{\text{amb}}}{R_{\text{lining-amb}}}
\end{align*}
where $T_{\text{lining},t}$ is the lining temperature, $R_{\text{M-lining}}$ is the thermal resistance between the metal pad and lining, $R_{\text{lining-amb}}$ is the lining-to-ambient thermal resistance, and $T_{\text{amb}}$ is the ambient temperature. The lining temperature then evolves as
\begin{equation}
C_{\text{lining}} \frac{d T_{\text{lining},t}}{d t} = Q_{\text{cathode},t} - Q_{\text{lining-amb},t}
\label{eq:lining_dynamics}
\end{equation}
where $C_{\text{lining}}$ is the thermal capacitance of the lining.

\textbf{4) Ledge-Coupled Sidewall Heat Transfer:} The sidewall heat loss pathway is particularly 
important because the phase-change process at the electrolyte-ledge interface pins the 
boundary temperature at the liquidus point $T_{\text{liq},t}$, thermally decoupling the 
electrolyte from the outer sidewall structure. Consequently, the heat flux extracted from the 
electrolyte depends only on the superheat and the bath-to-interface thermal resistance $R_
{\text{E-ledge}}$ \cite{taylor_energy_2014}:
\begin{equation}
Q_{\text{ledge},t} = \frac{T_{\text{super},t}}{R_{\text{E-ledge}}}
\label{eq:ledge_heat_transfer}
\end{equation}
The superheat is defined as the temperature difference between the bulk electrolyte and the liquidus temperature:
\begin{equation}
T_{\text{super},t} = T_{\text{E},t} - T_{\text{liq},t}
\label{eq:superheat_definition}
\end{equation}
where $T_{\text{liq},t}$ is the liquidus temperature, given by an empirical relation in $C_{\text{Al}_2\text{O}_3,t}$, $C_{\text{AlF}_3,t}$, $C_{\text{CaF}_2,t}$, and $C_{\text{MgF}_2,t}$ following Solheim \emph{et al.}~\cite{solheim_liquidus_2016}:
\begin{equation}
T_{\text{liq},t} = f_{\text{liq}}\!\left(C_{\text{Al}_2\text{O}_3,t},\,C_{\text{AlF}_3,t},\,C_{\text{CaF}_2,t},\,C_{\text{MgF}_2,t}\right).
\label{eq:liquidus_temp}
\end{equation}
The outward heat release from the ledge to the ambient is
\begin{equation}
Q_{\text{ledge-amb},t} = \frac{T_{\text{liq},t} - T_{\text{amb}}}{R_{\text{ledge-amb},t}}
\label{eq:ledge_amb_heat_loss}
\end{equation}
where the sidewall thermal resistance $R_{\text{ledge-amb},t}$ aggregates the ledge conduction, shell, and shell-to-ambient terms with a ledge-thickness dependence \cite{taylor_energy_2014}.

\subsection{Mass Balance Dynamics}
\label{sec:mass_balance}

Mass balance is required because the liquidus temperature depends on electrolyte composition. Over the scheduling horizon, the additions of $\text{AlF}_3$, $\text{CaF}_2$, $\text{MgF}_2$, and $\text{LiF}$ are neglected, so their concentrations are treated as known slow variables, while alumina is modeled explicitly because it changes with both feeding and electrolytic consumption \cite{wong_studies_2023}. This subsection therefore tracks the electrolyte mass, ledge thickness, and dissolved alumina concentration through the coupled effects of feeding, electrolysis, and ledge phase change.

{\color{blue}\textbf{1) Scheduling-Level Alumina-Feed Strategy:} To close these balances, the model specifies a scheduling-level alumina-feed strategy. In industrial practice, alumina feeding follows pseudo-resistance-based point-feeding or demand-feeding logic to maintain the dissolved concentration \cite{Robilliard2016,lavoie_review_2016}. At the hourly scheduling resolution, these minute-level feeding fluctuations are represented by a nominal average feed rate that tracks the current-dependent electrolytic consumption, thereby avoiding large concentration deviations caused by a fixed feed rate during power modulation. Because the state-dependent current efficiency $\eta_t$ depends on internal states predicted by the thermal-mass balance and is not directly available when a feed command is issued, this nominal rate uses the rated current efficiency $\eta_N$:}
\begin{equation}
\dot{m}_{\text{Al}_2\text{O}_3,t}^{\text{feed}} = \frac{M_{\text{Al}_2\text{O}_3}}{2 F z_{\text{Al}}} \cdot I_t \cdot \eta_N
\label{eq:feed_reference}
\end{equation}
{\color{blue}The alumina mass-balance equation propagates the difference between the nominal feed rate and actual electrolytic consumption into a predicted concentration trajectory. The scheduling model explicitly constrains this state trajectory within its upper and lower concentration bounds; in practice, minute-level cell controllers further correct feeding actions using pseudo-resistance and other concentration-related signals.}

\textbf{2) Electrolyte Mass and Ledge Thickness Dynamics:} The total molten electrolyte mass changes because of alumina feeding and mass exchange with the ledge:
\begin{equation}
\frac{d m_{\text{E}, t}}{d t} = \dot{m}_{\text{Al}_2\text{O}_3, t}^{\text{feed}} + \dot{m}_{\text{E}, t}^{\text{ledge}}
\label{eq:total_mass_balance}
\end{equation}
where $\dot{m}_{\text{Al}_2\text{O}_3, t}^{\text{feed}}$ is {\color{blue}the nominal alumina feed rate specified by \eqref{eq:feed_reference}}, and $\dot{m}_{\text{E}, t}^{\text{ledge}}$ is the mass exchange rate between molten electrolyte and ledge.

The ledge mass exchange rate follows the phase-change energy balance. The difference between $Q_{\text{ledge}, t}$ and $Q_{\text{ledge-amb},t}$ is absorbed or released as latent heat while maintaining the interface at $T_{\text{liq},t}$:
\begin{equation}
\dot{m}_{\text{E}, t}^{\text{ledge}} = \frac{Q_{\text{ledge}, t} - Q_{\text{ledge-amb},t}}{L_f}
\label{eq:ledge_mass_exchange}
\end{equation}
where positive values indicate ledge melting and negative values indicate ledge freezing. Assuming a uniform ledge profile over the area $A_{\text{ledge}}$, the ledge thickness evolves as
\begin{equation}
\frac{d L_{\text{ledge}, t}}{d t} = -\frac{1}{\rho_{\text{ledge}} \cdot A_{\text{ledge}}} \cdot \dot{m}_{\text{E}, t}^{\text{ledge}}
\label{eq:ledge_thickness_change}
\end{equation}
where $\rho_{\text{ledge}}$ is the ledge density.

\textbf{3) Alumina Mass and Concentration Balance:} The dissolved alumina mass evolves from the balance between feeding and electrolytic consumption:
\begin{equation}
\frac{d m_{\text{Al}_2\text{O}_3, t}}{d t} = \dot{m}_{\text{Al}_2\text{O}_3, t}^{\text{feed}} - \dot{m}_{\text{Al}_2\text{O}_3, t}^{\text{cons}}
\label{eq:alumina_mass_balance}
\end{equation}
where {\color{blue}$\dot{m}_{\text{Al}_2\text{O}_3, t}^{\text{feed}}$ is the nominal scheduling-level alumina feed rate specified by \eqref{eq:feed_reference}, and $\dot{m}_{\text{Al}_2\text{O}_3, t}^{\text{cons}}$ is the electrolytic consumption rate.} The alumina consumption rate is given by Faraday's law with the state-dependent current efficiency $\eta_t$:
\begin{equation}
\dot{m}_{\text{Al}_2\text{O}_3, t}^{\text{cons}} = \frac{M_{\text{Al}_2\text{O}_3}}{2 F z_{\text{Al}}} \cdot I_t \cdot \eta_t
\label{eq:alumina_consumption}
\end{equation}

Because the alumina dissolution time constant is much shorter than the hourly scheduling horizon, dissolution kinetics are neglected, and the fed alumina is assumed to dissolve instantaneously \cite{lavoie_review_2016}. The dissolved alumina concentration in wt.\% then evolves as
\begin{equation}
\frac{d C_{\text{Al}_2\text{O}_3, t}}{d t} = \frac{100}{m_{\text{E}, t}} \left( \frac{d m_{\text{Al}_2\text{O}_3, t}}{d t} - \frac{C_{\text{Al}_2\text{O}_3, t}}{100} \frac{d m_{\text{E}, t}}{d t} \right)
\label{eq:concentration_change}
\end{equation}
where $C_{\text{Al}_2\text{O}_3, t} = 100\,m_{\text{Al}_2\text{O}_3, t}/m_{\text{E}, t}$.

\section{Computationally Tractable Reformulation and Solution Methodology}
\label{sec:reformulation}
The TMC dynamic model in Section \ref{sec:cell_model} contains multiple nonlinear elements, including quadratic terms, nonlinear empirical state functions, and bilinear products. Directly embedding this nonlinear state-space model into the grid scheduling framework (Section II) leads to a Mixed-Integer Nonlinear Programming (MINLP) problem that is generally non-convex and computationally demanding for large-scale or real-time applications. This section therefore reformulates the model into a solver-ready MILP using a set of linearization steps.

\textbf{1) Power-current quadratic term:}
As defined in \eqref{eq:power} and \eqref{eq:voltage}, the cell power consumption $P_t$ exhibits a quadratic dependence on the load current $I_t$ because of the ohmic voltage drop. Since the current modulation range considered here is narrow relative to the rated current, this relation is linearized locally through a first-order Taylor expansion around $I_N$:
\begin{equation}
P_t = P_N + (V_{\text{EMF}} + 2 \cdot R_{\text{ohm}} \cdot I_N) \cdot (I_t - I_N)
\label{eq:power_linearization_s4}
\end{equation}
where $I_N$ denotes the rated current of the representative cell and $P_N$ denotes the corresponding rated power:
\begin{equation}
P_N = (V_{\text{EMF}} + R_{\text{ohm}} \cdot I_N) \cdot I_N.
\label{eq:rated_power_s4}
\end{equation}

{\color{blue}As a local representation, the approximation error increases as the current moves away from $I_N$. If a wider current range must be represented, the power--current relation can be refined using a piecewise-linear approximation.}

\textbf{2) Empirical state-function fitting:}
The original empirical models for current efficiency $\eta_t$ \eqref{eq:current_efficiency} and liquidus temperature $T_{\text{liq},t}$ \eqref{eq:liquidus_temp} are highly nonlinear, involving logarithmic, exponential, and higher-order polynomial terms. Within the operating range considered here, however, the relevant state variables remain confined to a narrow feasible region by the safety constraints in Section II. Over this local region, multi-variable linear regressions provide an adequate approximation:
\begin{equation}
T_{\text{liq},t} = \beta_1 \cdot C_{\text{Al}_2\text{O}_3,t} + \beta_2 \cdot C_{\text{AlF}_3,t} + \beta_3 \cdot C_{\text{CaF}_2,t} + \beta_4 \cdot C_{\text{MgF}_2,t} + \beta_5
\label{eq:liquidus_linear_s4}
\end{equation}
\begin{equation}
\eta_t = \alpha_1 \cdot T_{\text{super},t} + \alpha_2 \cdot C_{\text{AlF}_3,t} + \alpha_3 \cdot C_{\text{LiF},t} + \alpha_4.
\label{eq:efficiency_linear_s4}
\end{equation}

\textbf{3) McCormick envelope for bilinear products:}
In the proposed TMC model, multiple bilinear coupled terms emerge naturally from the physical laws. Specifically, the aluminum production rate \eqref{eq:production} introduces the product of current efficiency and load current, while the mass-concentration dynamics \eqref{eq:concentration_change} involve the product of species concentration and total electrolyte mass.
For a bilinear term $Z = X \cdot Y$ with $X \in [X^{\min}, X^{\max}]$ and $Y \in [Y^{\min}, Y^{\max}]$, the McCormick envelope is:
\begin{align}
Z &\geq X^{\min} \cdot Y + Y^{\min} \cdot X - X^{\min} \cdot Y^{\min} \\
Z &\geq X^{\max} \cdot Y + Y^{\max} \cdot X - X^{\max} \cdot Y^{\max} \\
Z &\leq X^{\min} \cdot Y + Y^{\max} \cdot X - X^{\min} \cdot Y^{\max} \\
Z &\leq X^{\max} \cdot Y + Y^{\min} \cdot X - X^{\max} \cdot Y^{\min}
\label{eq:mccormick_s4}
\end{align}
{\color{blue}The quality of the McCormick envelope relaxation depends on the tightness of the variable bounds. In this application, the physical safety constraints restrict the relevant state variables and thereby reduce the relaxation gap in the normal operating region. Where tighter accuracy is required, selected variable boxes can be partitioned to refine the envelope.}

{\color{blue}After these reformulations, the scheduling model becomes a MILP that can be solved efficiently with commercial solvers.}

\section{Case Study}
\label{sec:case_study}

\subsection{Case Study Setup}
\label{sec:case_setup}

    The case study considers a representative 500\,kA potline with 280 electrolytic cells. {\color{blue}This corresponds to a rated potline power of approximately 553\,MW, a scale consistent with modern high-amperage aluminum-smelting technologies \cite{taylor_energy_2014}. The framework is evaluated in two illustrative demand response scenarios. The PBDR case uses a 24-hour horizon with $\Delta t=1\,\mathrm{h}$ and the known Shandong ToU tariff, i.e., $\pi_t^{\mathrm e}=\pi_t^{\mathrm{ToU}}$ \cite{bjx_shandong_tou_2025}. The IBDR case uses an 8-hour event-and-recovery window with a mandatory 2-hour event announced 1\,h in advance; $P_t^{\mathrm{base}}=P_N$, $P_{\text{target},t}=0.85P_N$, and $\pi_t^{\mathrm{DR}}=4\pi_t^{\mathrm{ToU}}$ during the event. The main safety limits are set to 950--970\,$^\circ$C for electrolyte temperature and 5--15\,$^\circ$C for superheat, and the effective ledge-thickness state is constrained to 0.08--0.14\,m, representing typical operating windows and model bounds \cite{reek_power_2015,wong_studies_2023}. The thermal and ledge parameters are representative equivalent lumped parameters selected from cell geometry, physical-property assumptions, and aluminum-electrolysis thermal-mass balance relations \cite{wong_studies_2023}. The corresponding numerical settings are summarized in Table~\ref{tab:case_params}.} The terminal recovery constraint is enforced on the electrolyte temperature ($\epsilon = 2\,^\circ\mathrm{C}$) and ledge thickness ($\epsilon = 0.01\,\mathrm{m}$).

\begin{table*}[!t]
\renewcommand{\arraystretch}{1.08}
\caption{\color{blue}Key Case-Study Parameters.}
\label{tab:case_params}
\centering
\scriptsize\color{blue}
\begin{tabular}{@{}p{0.13\textwidth}p{0.28\textwidth}p{0.54\textwidth}@{}}
\toprule
Category & Parameters & Case-study setting \\
\midrule
Potline scale & Rated current; number of cells; rated potline power & 500\,kA; 280; approximately 553\,MW \\
Scheduling setup & Horizon; time step; current range & 24\,h; 1\,h; 350--550\,kA \\
Safety limits & Electrolyte temperature; superheat; alumina concentration; effective ledge thickness & 950--970\,$^\circ$C; 5--15\,$^\circ$C; 2.5--3.5\,wt.\%; 0.08--0.14\,m \\
Initial state & Electrolyte temperature; liquidus temperature; superheat; alumina concentration; effective ledge thickness & 966\,$^\circ$C; 953.67\,$^\circ$C; 12.33\,$^\circ$C; 3.0\,wt.\%; 0.09\,m \\
Thermal model & $R_{\mathrm{E-gas}}$; $R_{\mathrm{E-ledge}}$; $R_{\mathrm{M-lining}}$; $R_{\mathrm{lining-amb}}$; $C_{\mathrm{lining}}$; $m_{\mathrm{M}}c_{\mathrm{M}}$; $m_{\mathrm{E,init}}$ & $1.762{\times}10^{-3}$, $4.350{\times}10^{-5}$, $2.017{\times}10^{-3}$, $4.585{\times}10^{-3}$\,K/W; $9.25{\times}10^7$\,J/K; $4.45{\times}10^7$\,J/K; $9.324{\times}10^3$\,kg \\
Ledge model & $\rho_{\mathrm{ledge}}$; $L_f$; $k_{\mathrm{ledge}}$; $b_{\mathrm{ledge}}$; $\rho_{\mathrm{ledge}}A_{\mathrm{ledge}}$ & 3000\,kg/m$^3$; $3.10{\times}10^5$\,J/kg; $3.307{\times}10^{-2}$\,K\,m/W; $2.813{\times}10^{-4}$\,K/W; $6.048{\times}10^4$\,kg/m \\
Electrochemical model & Back EMF; cell resistance; rated current efficiency; efficiency bounds & 2.16\,V; $3.58{\times}10^{-3}$\,m$\Omega$; 0.93; 0.91--0.95 \\
Composition and environment & Initial Al$_2$O$_3$, AlF$_3$, CaF$_2$, LiF, MgF$_2$; ambient/gas temperature & 3.0, 11.0, 6.0, 1.5, 1.0\,wt.\%; 30/50\,$^\circ$C \\
\bottomrule
\end{tabular}
\par\vspace{0.5mm}
\end{table*}

To systematically evaluate the contribution of different modeling components, three models with increasing physical fidelity are compared:

\begin{itemize}
    \item \textbf{Linear:} A steady-state model that assumes the electrolyte temperature varies proportionally with the input power, thereby neglecting all heat-dissipation dynamics \cite{yue_collaborative_2024}.
    \item \textbf{ETP:} An Equivalent Thermal Parameter model that introduces a first-order RC thermal circuit to capture the electrolyte temperature dynamics \cite{huang_novel_2024}. However, it treats the ledge thickness as static and omits the mass balance subsystem (Section~\ref{sec:mass_balance}), thereby eliminating the phase-change interface and lumping the entire sidewall into a single equivalent thermal resistance.
    \item \textbf{TMC (proposed):} The complete Thermal-Mass Coupled framework (Sections~\ref{sec:cell_model} and \ref{sec:reformulation}), which explicitly captures the coupled thermal-mass dynamics including ledge phase-change dynamics and state-dependent current efficiency.
\end{itemize}
{\color{blue}For a fair comparison, the three models use the same scheduling-level alumina-feed strategy \eqref{eq:feed_reference} and are parameterized from the same physical parameter set.}

{\color{blue}To assess the effect of model approximation on schedule feasibility, a nonlinear reference TMC model is used as a common evaluator. This reference model is a direct numerical implementation of Section~\ref{sec:cell_model} that retains the nonlinear relations before the MILP reformulation in Section~\ref{sec:reformulation}. The state trajectory generated internally by each optimization model is referred to as the \textit{predicted} trajectory, whereas the trajectory obtained by applying the same current schedule to the nonlinear reference model is referred to as the \textit{reference} trajectory. This comparison evaluates the models rather than validates them against plant measurements. Validation against measurements from instrumented industrial cells is left for future work.}

All optimization problems are formulated as MILPs and solved using MATLAB (R2025b) \cite{matlab} with YALMIP \cite{lofberg_yalmip_2004} as the modeling interface and {\color{blue}Gurobi (V12.0.3)} \cite{gurobi} as the solver. {\color{blue}The relative MIP optimality gap is set to $10^{-3}$ for the PBDR cases and $10^{-4}$ for the IBDR cases.} Computations are performed on an Apple M4 processor with 16\,GB of unified memory. {\color{blue}All MILP instances are solved to the specified optimality gap within 10 seconds, well below the hourly scheduling interval of the illustrative cases.}

\subsection{Power Modulation Impact on Temperature}

\begin{figure}[t]
  \centering
  \includegraphics[width=\columnwidth]{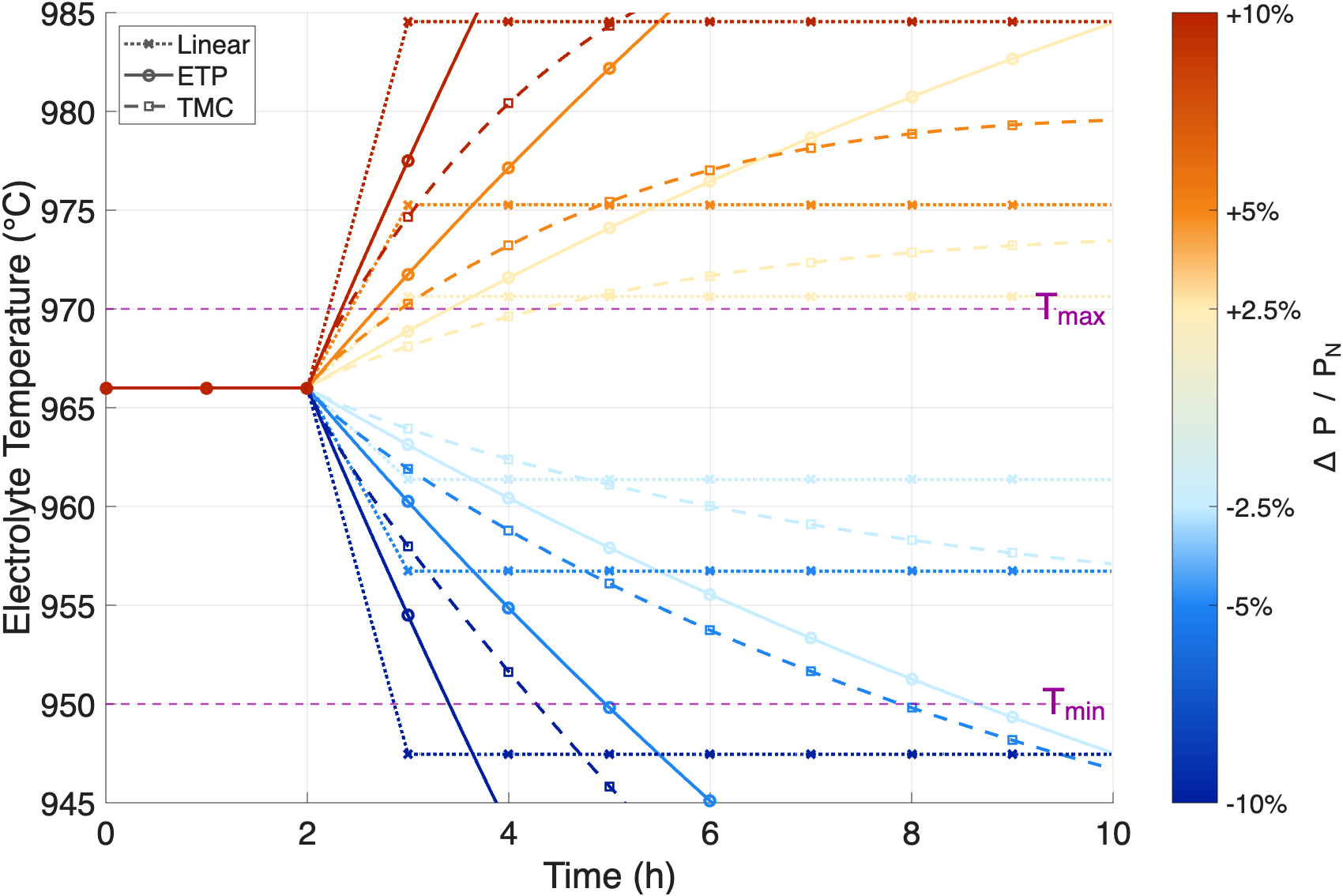}
  \caption{Open-loop electrolyte-temperature trajectories of the Linear, ETP, and TMC models under $\pm 2.5\%$, $\pm 5\%$, and $\pm 10\%$ step changes in potline power applied at $t = 2$\,h. Dashed lines mark the safety bounds $T_{\min} = 950\,^\circ\mathrm{C}$ and $T_{\max} = 970\,^\circ\mathrm{C}$.}
  \label{fig:1}
\end{figure}

Fig.~\ref{fig:1} compares the open-loop electrolyte-temperature response of the three models to step changes in potline power, with each curve obtained from direct nonlinear simulation of the corresponding model. The distinct trajectory shapes follow directly from the model structures. The Linear model maps $(P_t - P_{t-1})\,\Delta t$ directly to the change in $T_{\text{E},t}$ without any explicit heat-loss term and therefore has no thermal time constant; all trajectories jump at $t = 2$\,h and then remain flat. The ETP model adds lumped heat loss and therefore produces a first-order exponential response, but because the side ledge is fixed its effective thermal inertia is too small, causing the temperature to cross the {\color{blue}safety limits} faster than in the TMC model. The TMC model further couples \eqref{eq:thermal_balance}, \eqref{eq:ledge_heat_transfer}, and \eqref{eq:ledge_mass_exchange}, so ledge phase change introduces a slower buffering timescale. Under load reduction, latent-heat release and reduced sidewall heat loss delay the temperature drop. As a result, the $-5\%$ case reaches $T_{\min}$ near 5\,h in ETP and only around 8\,h in TMC. This explains why accurate multi-hour thermal dynamics matter for the scheduling results in Figs.~\ref{fig:2} and \ref{fig:3}.

\subsection{State Variable Comparison}

\begin{figure*}[!t]
  \centering
  \includegraphics[width=\textwidth]{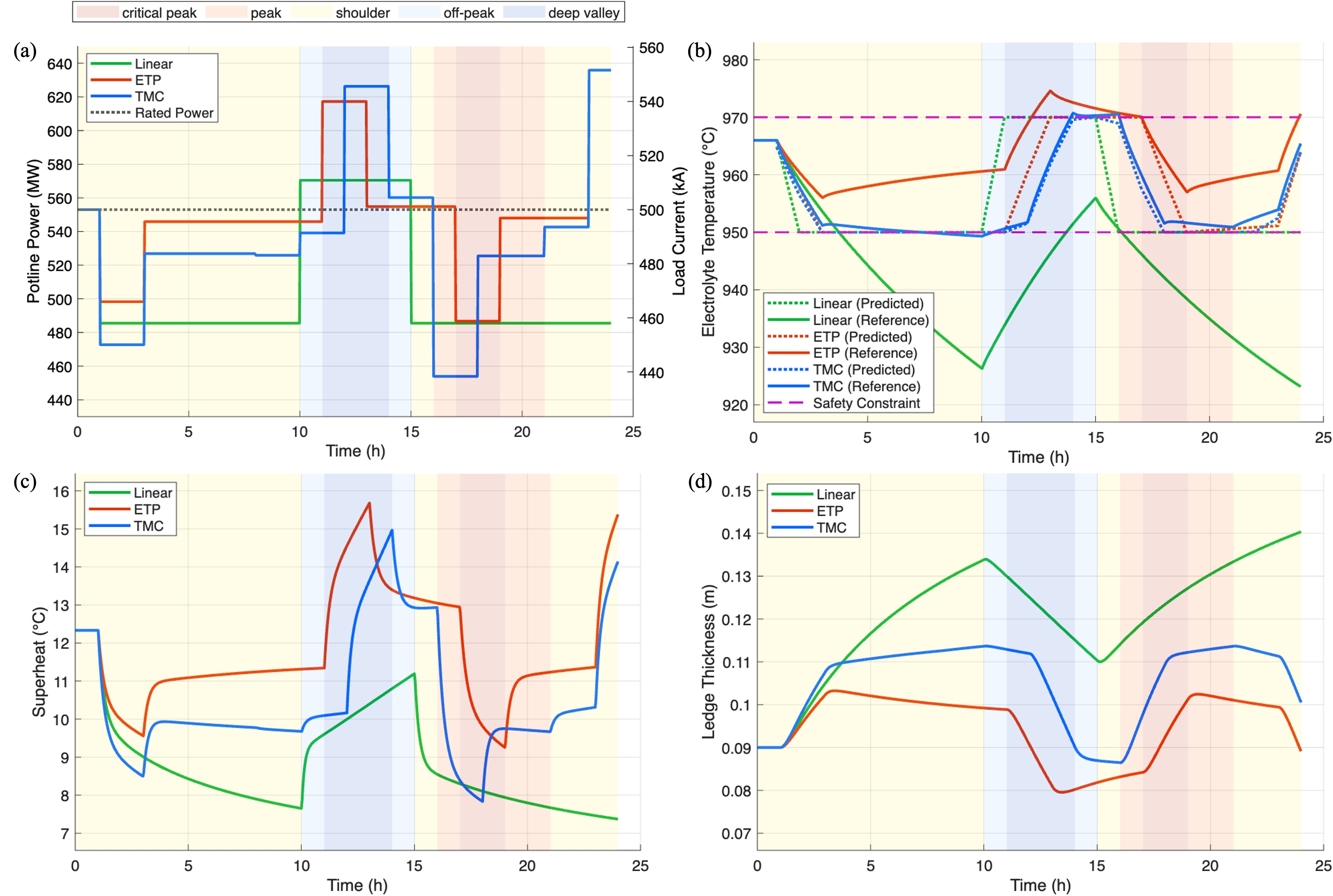}
  \caption{Scheduled power and state trajectories in the {\color{blue}PBDR} scenario: (a) potline power{\color{blue}, rated power}, and load current; (b) electrolyte temperature, with optimizer-predicted (dashed) and reference-model (solid) trajectories; (c) superheat; and (d) ledge thickness.}
  \label{fig:2}
\end{figure*}

\begin{figure*}[!t]
  \centering
  \includegraphics[width=\textwidth]{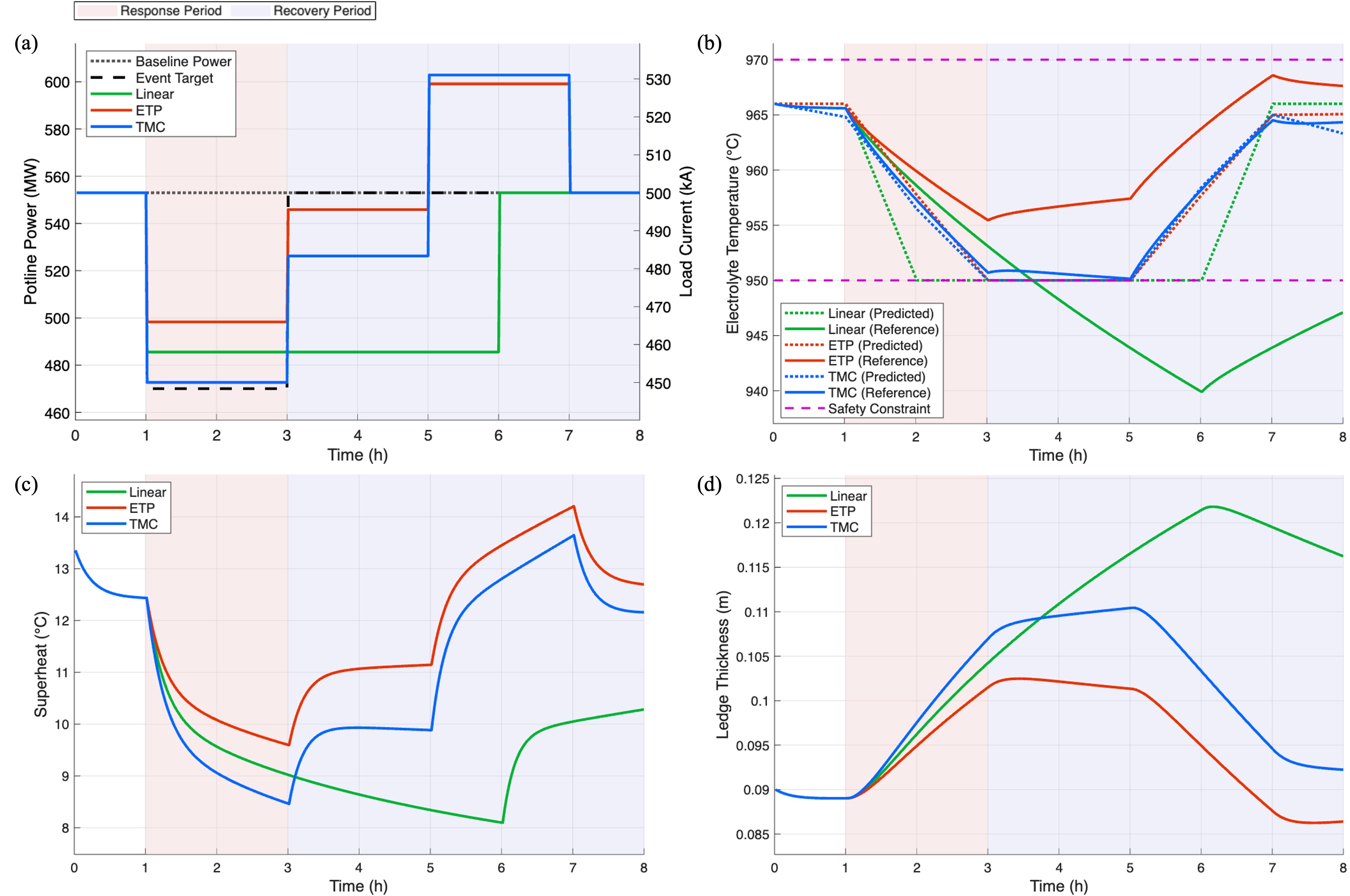}
  \caption{Scheduled power and state trajectories in the {\color{blue}IBDR} scenario: (a) potline power{\color{blue}, settlement baseline, event target}, and load current; (b) electrolyte temperature, with optimizer-predicted (dashed) and reference-model (solid) trajectories; (c) superheat; and (d) ledge thickness.}
  \label{fig:3}
\end{figure*}

In Figs.~\ref{fig:2} and \ref{fig:3}, subplot (a) reports the scheduled potline power and load current; (b) compares the \textit{predicted} (dashed) and \textit{reference} (solid) electrolyte temperatures, the dominant active safety constraint; (c) and (d) show the reference superheat and ledge-thickness trajectories that drive the thermal response in (b).

{\color{blue}\textbf{1) Price-Based Mechanism Analysis (Fig.~\ref{fig:2}):}}
The PBDR results reveal two different failure modes of the simplified models. The ETP model is overly conservative because it underestimates the feasible load-reduction margin during the initial flat-price period. At Hours 2--3, it lowers power only to about 500\,MW and then restores the load close to rated operation by Hour 4, since its \textit{predicted} temperature trajectory has already approached the lower bound. However, the reference trajectory in Fig.~\ref{fig:2}(b) shows that {\color{blue}the electrolyte-temperature trajectory from the nonlinear reference model} remains several degrees above this predicted bound, indicating that the cell still retains exploitable thermal margin after the early load reduction. Because this margin is not fully used in the flat-price period, the later valley-price period (Hours 11--15) no longer has sufficient room for upward regulation. As a result, when power is increased at Hours 12--13, the reference electrolyte temperature exceeds the upper safety limit.

The Linear model makes the opposite error by scheduling overly deep power reductions. By assuming an almost instantaneous proportional relation between power and temperature, it keeps suppressing power whenever the electricity price is high, while failing to capture the cumulative cooling effect caused by thermal inertia. As a result, although the optimizer's predicted trajectory satisfies its own temperature constraint, {\color{blue}the electrolyte-temperature trajectory from the nonlinear reference model} falls well below the lower safety bound, indicating a substantial risk of electrolyte freeze-out.

By contrast, the TMC model achieves a physically consistent balance between flexibility and safety. After the initial regulation, it returns to an intermediate level of about 525\,MW and can still reduce power to about 455\,MW in the later high-price interval while keeping {\color{blue}the electrolyte-temperature trajectory from the nonlinear reference model close to the prescribed temperature limits}. The mechanism is visible in Figs.~\ref{fig:2}(c) and \ref{fig:2}(d): during the later load reduction, the superheat drops from about $13\,^\circ\mathrm{C}$ to $8\,^\circ\mathrm{C}$, while the ledge thickens from about 0.087\,m to 0.112\,m as the interface solidifies. The lower superheat weakens sidewall heat loss, and the thicker ledge further increases the thermal buffer. These coupled state changes slow the temperature decline and preserve {\color{blue}thermal margin}.

{\color{blue}\textbf{2) Incentive-Based Mechanism Analysis (Fig.~\ref{fig:3}):}}
{\color{blue}The IBDR scenario imposes a stricter requirement than PBDR because the ASL must closely follow the notified target while still satisfying all state constraints and the terminal recovery constraint.}

The main distinction in this scenario is dispatch accuracy. The ETP model again remains too conservative. Once its internal temperature prediction approaches the lower bound, it refuses further load reduction and therefore falls short of the {\color{blue}notified target}. In Fig.~\ref{fig:3}(a), it reduces power only to about 500\,MW, whereas the target is about 470\,MW. The reference trajectory in Fig.~\ref{fig:3}(b), however, shows that {\color{blue}the electrolyte-temperature trajectory from the nonlinear reference model} stays several degrees above the predicted bound, confirming that the ETP model overstates the thermal constraint because it does not capture the ledge-buffering effect discussed above.

By contrast, the TMC model follows the dispatch instruction much more closely. In Fig.~\ref{fig:3}(a), it tracks the notified target near 470\,MW, corresponding to a power reduction of about 85\,MW from rated operation, while the predicted and reference electrolyte-temperature trajectories remain closely aligned within {\color{blue}the prescribed $950$--$970\,^\circ\mathrm{C}$ temperature limits}. After the DR event, the cell states return smoothly toward their initial levels without overshoot, showing that accurate dispatch tracking and cyclic-state recovery can be achieved simultaneously.

\subsection{Economic Comparison}

{\color{blue}\textbf{1) Base-Case Economic Comparison:}} Tables \ref{tab:3} and \ref{tab:4} compare the economic performance of the three models in the {\color{blue}PBDR} and {\color{blue}IBDR} scenarios, respectively. All entries are measured against constant rated-power operation and reported in nominal 2025 CNY ($1\,\mathrm{USD} \approx 7.2\,\mathrm{CNY}$). The Linear model is included for comparison, with a \emph{Violation Penalty} column that monetizes safety-bound violations via two mechanisms. Throughout, $T_{\text{ref}}(t)$ denotes the reference-model electrolyte temperature defined in Section~\ref{sec:case_setup}. First, when the cumulative undertemperature severity $S_{\text{low}} = \int_0^T \max(T_{\min}-T_{\text{ref}}(t),\,0)\,\mathrm{d}t$ exceeds a threshold $\tau_{\text{freeze}} = 5\,^\circ\mathrm{C}\cdot\mathrm{h}$ (corresponding to the 3--4\,h significant-cooling regime of high-amperage cells \cite{chesonis_lengthy_2019}), the cell is booked as an unrecoverable freeze-out at a cost $C_{\text{freeze}} = 3\times 10^6$\,CNY, a conservative cell-level lower bound well below reported potline-level restart costs \cite{chesonis_lengthy_2019}. Second, the over-temperature severity $S_{\text{high}} = \int_0^T \max(T_{\text{ref}}(t)-T_{\max},\,0)\,\mathrm{d}t$ is penalized at $C_{\text{deg}} = 2{,}000$\,CNY/$(^\circ\mathrm{C}\cdot\mathrm{h})$, taken here as a rough estimate of cathode-life cost under sustained overheating. Once a freeze-out is triggered, the Linear schedule's remaining ledger entries lose physical meaning and are reported as ``--''.

\begin{table*}[t]
  \begin{minipage}{0.48\textwidth}
    \caption{Economic comparison in the {\color{blue}PBDR} scenario.}
    \label{tab:3}
    \centering
    \resizebox{\linewidth}{!}{%
    \begin{tabular}{l c c c c c}
    \toprule
    \textbf{Model} & \textbf{\makecell{Elec.\\Saving}} & \textbf{\makecell{Alumina\\Saving}} & \textbf{\makecell{Yield\\Penalty}} & \textbf{\makecell{Violation\\Penalty}} & \textbf{\makecell{Net\\Improvement}} \\
    \midrule
    Linear & -- & -- & -- & $-3{,}000{,}000$ & $-3{,}000{,}000$ \\
    ETP & 226,899 & 36,918 & -75,226 & -17,797 & 170,794 \\
    TMC & {\color{blue}575,053} & {\color{blue}145,322} & {\color{blue}-339,811} & {\color{blue}-671} & {\color{blue}379,892} \\
    \bottomrule
    \end{tabular}%
    }
  \end{minipage}\hfill
  \begin{minipage}{0.50\textwidth}
    \caption{Economic comparison in the {\color{blue}IBDR} scenario.}
    \label{tab:4}
    \centering
    \resizebox{\linewidth}{!}{%
    \begin{tabular}{l c c c c c c}
    \toprule
    \textbf{Model} & \textbf{\makecell{Elec.\\Saving}} & \textbf{\makecell{Alumina\\Saving}} & \textbf{\makecell{Yield\\Penalty}} & \textbf{\makecell{DR\\Revenue}} & \textbf{\makecell{Violation\\Penalty}} & \textbf{\makecell{Net\\Improvement}} \\
    \midrule
    Linear & -- & -- & -- & -- & $-3{,}000{,}000$ & $-3{,}000{,}000$ \\
    ETP & 67,826 & 11,687 & -23,822 & 179,789 & 0 & 235,479 \\
    TMC & {\color{blue}183,485} & 40,186 & {\color{blue}-102,747} & 880,148 & 0 & {\color{blue}1,001,072} \\
    \bottomrule
    \end{tabular}%
    }
  \end{minipage}
\end{table*}

The economic results reflect the state-trajectory differences discussed above. In the {\color{blue}PBDR} scenario, the net economic improvement achieved by the TMC model is more than twice that of the ETP model, increasing from 170,794\,CNY to {\color{blue}379,892}\,CNY (Table~\ref{tab:3}). The main contributor is the much larger electricity-cost saving, which rises from 226,899\,CNY to {\color{blue}575,053}\,CNY because the TMC model can exploit price differences with deeper load shifting. The alumina-cost saving also increases markedly, from 36,918\,CNY to {\color{blue}145,322}\,CNY. Although the TMC schedule incurs a larger yield penalty, that penalty is more than compensated by the additional arbitrage value enabled by the wider feasible operating envelope.

In the {\color{blue}IBDR} scenario, the advantage is more pronounced, with the net economic improvement increasing from 235,479\,CNY to {\color{blue}1,001,072}\,CNY (Table~\ref{tab:4}). The dominant source is the DR settlement. Because the TMC model can track the {\color{blue}notified target} closely, DR revenue rises from 179,789\,CNY under ETP to 880,148\,CNY under TMC; the ETP model forgoes most of this revenue by shedding load too cautiously. The TMC model also improves the electricity-cost saving and the alumina-cost saving, although it again incurs a larger yield penalty. {\color{blue}Across both scenarios, the economic gains reflect the additional flexibility identified by the TMC model.}

The Linear rows illustrate the safety consequence of neglecting thermal dynamics. As defined above, a cell is classified as an unrecoverable freeze-out when its cumulative undertemperature severity $S_{\text{low}}$ exceeds the threshold $\tau_{\text{freeze}} = 5\,^\circ\mathrm{C}\cdot\mathrm{h}$. The Linear schedules reach $S_{\text{low}} = 236.2\,^\circ\mathrm{C}\cdot\mathrm{h}$ in PBDR and $25\,^\circ\mathrm{C}\cdot\mathrm{h}$ in IBDR, far exceeding this threshold and triggering the freeze-out penalty in both cases. By contrast, the TMC schedules keep $S_{\text{low}} \leq 1\,^\circ\mathrm{C}\cdot\mathrm{h}$ in both scenarios, {\color{blue}well below the freeze-out threshold}. {\color{blue}This contrast confirms that accurate thermal-mass coupling is necessary to reduce freeze-out risk while improving economic scheduling performance.}

{\color{blue}\textbf{2) Sensitivity Analysis:} To further examine the dependence of the economic conclusions on market-side assumptions, additional sensitivity cases are solved and then evaluated using the same nonlinear reference-model assessment and economic accounting as in Tables~\ref{tab:3} and \ref{tab:4}. For IBDR, the baseline load is varied as $P_t^{\mathrm{base}}=\kappa P_N$, with $\kappa\in\{0.95,1.00,1.05\}$. The main case takes the rated potline power as the baseline load, and the scheduled load returns close to the normal operating level after the DR event to recover the thermal-mass states. Therefore, the $\pm5\%$ range represents a moderate uncertainty in the baseline load around this normal operating level, rather than the full technical power-modulation range. The target response is kept at 15\% of the corresponding baseline load, so both the credited response and the response-performance ratio change with $\kappa$. Here, the mean response ratio is defined as $\bar{\gamma}=|\mathcal{T}_{\text{mandatory}}|^{-1}\sum_{t\in\mathcal{T}_{\text{mandatory}}}\gamma_t$. A value close to one indicates close target tracking, whereas values in the range $0.8\leq\gamma_t<1.2$ receive the full compensation multiplier. As shown in Table~\ref{tab:baseline_sensitivity}, the absolute IBDR value is highly sensitive to the baseline load. A lower baseline load sharply reduces the credited response and pushes the ETP schedule below the compensated response range, whereas the TMC schedule still earns partial DR revenue. Across all tested baseline-load levels, the TMC schedule achieves a net economic advantage over ETP, ranging from 237.9 to 765.6\,kCNY.}

\begin{table}[!t]
{\color{blue}
\renewcommand{\arraystretch}{1.00}
\caption{IBDR Baseline-Load Sensitivity Based on Nonlinear Reference-Model Assessment.}
\label{tab:baseline_sensitivity}
\centering
\scriptsize
\setlength{\tabcolsep}{2.2pt}
\begin{tabular}{c cc cc cc}
\toprule
\makecell{Baseline\\factor}
& \multicolumn{2}{c}{\makecell{Mean response\\ratio $\bar{\gamma}$}}
& \multicolumn{2}{c}{\makecell{DR revenue}}
& \multicolumn{2}{c}{\makecell{Net improvement}} \\
\cmidrule(lr){2-3}\cmidrule(lr){4-5}\cmidrule(l){6-7}
& ETP & TMC & ETP & TMC & ETP & TMC \\
\midrule
0.95 & 0.343 & 0.668 & 0.0 & 172.9 & 55.7 & \makecell{293.6\\(+427.2\%)} \\
1.00 & 0.659 & 0.969 & 179.8 & 880.1 & 235.5 & \makecell{1001.1\\(+325.1\%)} \\
1.05 & 0.946 & 1.174 & 902.2 & 1119.9 & 957.9 & \makecell{1238.3\\(+29.3\%)} \\
\bottomrule
\end{tabular}
\par\vspace{0.5mm}
\noindent\parbox{\columnwidth}{\scriptsize\emph{Note:} DR revenue and net improvement are in $10^3$ CNY. Parenthetical percentages in the TMC net-improvement column are relative to ETP at the same baseline factor.}
}
\end{table}

{\color{blue}For PBDR, the ToU price spread is varied while preserving the daily average price and the peak/valley period structure, i.e., $\pi_t^{s}=\bar{\pi}+s(\pi_t-\bar{\pi})$ with $s\in\{0.8,1.0,1.2\}$. The violation-penalty coefficients $C_{\text{freeze}}$ and $C_{\text{deg}}$ are then scaled by 0.5--2.0 in the post-simulation assessment. Table~\ref{tab:tou_penalty_sensitivity} shows that a wider ToU spread increases the net value for both ETP and TMC, and the TMC advantage remains positive across all tested spreads. In addition, scaling the violation penalty changes the ETP net value by about 26.7--26.8\,kCNY in each ToU-spread case, whereas the TMC net value changes by less than 2.5\,kCNY. This difference occurs because the TMC schedules remain closer to the temperature limits in the nonlinear reference-model assessment, while the ETP schedules incur larger over-temperature penalties. Therefore, the model-comparison conclusion is not driven by a particular penalty coefficient.}

\begin{table*}[!t]
{\color{blue}
\renewcommand{\arraystretch}{0.96}
\caption{PBDR Sensitivity to ToU Spread and Violation-Penalty Coefficients.}
\label{tab:tou_penalty_sensitivity}
\centering
\footnotesize
\setlength{\tabcolsep}{2.2pt}
\begin{tabular*}{\textwidth}{@{\extracolsep{\fill}}c cc cc cc@{}}
\toprule
ToU factor
& \multicolumn{2}{c}{Penalty 0.5$\times$}
& \multicolumn{2}{c}{Penalty 1.0$\times$}
& \multicolumn{2}{c}{Penalty 2.0$\times$} \\
\cmidrule(lr){2-3}\cmidrule(lr){4-5}\cmidrule(l){6-7}
& ETP & TMC (+\%)
& ETP & TMC (+\%)
& ETP & TMC (+\%) \\
\midrule
0.8 & 152.7 & 341.8 (+123.9\%) & 143.8 & 341.3 (+137.3\%) & 126.0 & 340.2 (+170.0\%) \\
1.0 & 179.7 & 380.2 (+111.6\%) & 170.8 & 379.9 (+122.4\%) & 153.0 & 379.2 (+147.9\%) \\
1.2 & 207.3 & 413.9 (+99.6\%) & 198.5 & 413.3 (+108.2\%) & 180.7 & 412.1 (+128.0\%) \\
\bottomrule
\end{tabular*}
\noindent\parbox{\textwidth}{\scriptsize\emph{Note:} Net improvements are in $10^3$ CNY from nonlinear-reference assessments; parentheses give the TMC gain over ETP at the same setting.}
}
\end{table*}

\subsection{{\color{blue}Approximation-Error and Feasibility Assessment}}
\label{sec:approx_error_audit}

{\color{blue}For each positive nonlinear relation $y$ and its approximation $\hat{y}$, the normalized discrepancy is evaluated as}
{\color{blue}\begin{equation}
e_t = 100\frac{|\hat{y}_t-y_t|}{|y_t|}.
\label{eq:normalized_approx_error}
\end{equation}}

{\color{blue}Table~\ref{tab:approx_error} distinguishes deterministic domain-screening grids from discrepancies realized along the optimized TMC trajectories. The domain screening shows that a single Taylor expansion around $I_N$ should not be interpreted as uniformly accurate across the complete 350--550\,kA current interval. By contrast, the optimized PBDR and IBDR trajectories remain in narrower regions, where the Taylor discrepancies are 0.911\% and 0.566\%, respectively. The empirical linear-regression fits remain below 0.52\% over their prescribed fitting domains and below 0.12\% along both optimized trajectories. The maximum McCormick-envelope deviations along the optimized PBDR and IBDR trajectories are 5.155\% and 4.959\%, respectively. Overall, the realized discrepancies remain modest over the optimized trajectories.}

\begin{table}[!t]
{\color{blue}
\renewcommand{\arraystretch}{0.98}
\caption{Approximation Error Audit.}
\label{tab:approx_error}
\centering
\footnotesize
\setlength{\tabcolsep}{1.6pt}
\begin{tabular*}{\columnwidth}{@{\extracolsep{\fill}}p{0.39\columnwidth}ccc@{}}
\toprule
\makecell[l]{Approximation\\item} & \makecell[c]{Domain-screening\\grid} & \makecell[c]{PBDR\\trajectory} & \makecell[c]{IBDR\\trajectory} \\
\midrule
Power--current Taylor approximation & 6.743\% & 0.911\% & 0.566\% \\
Current-efficiency linear-regression fit & 0.514\% & 0.083\% & 0.083\% \\
Liquidus-temperature linear-regression fit & 0.284\% & 0.117\% & 0.117\% \\
McCormick bilinear products & 5.263\% & 5.155\% & 4.959\% \\
\bottomrule
\end{tabular*}
\noindent\parbox{\columnwidth}{\scriptsize\emph{Note:} All entries are maximum relative discrepancies. Domain screening uses deterministic uniform grids: 1001 Taylor points over 350--550\,kA; $51^3$ points for the current-efficiency linear-regression fit over 0--20\,$^\circ$C superheat, 9--13\,wt.\% AlF$_3$, and 1--2\,wt.\% LiF; $41{\times}41{\times}41{\times}21$ points for the liquidus-temperature linear-regression fit over 2--4\,wt.\% Al$_2$O$_3$, 8--15\,wt.\% AlF$_3$, 4--8\,wt.\% CaF$_2$, and 0.5--1.5\,wt.\% MgF$_2$; and $201{\times}201$ points for each McCormick variable box. Trajectory values are maxima over optimized hourly points.}
}
\end{table}

{\color{blue}Each optimized current trajectory is then applied to the nonlinear reference TMC model used in Figs.~\ref{fig:2} and \ref{fig:3} to assess the resulting schedules. In the PBDR case, the maximum hourly electrolyte-temperature difference between the MILP-predicted and reference trajectories is 2.41$\,^{\circ}\mathrm{C}$; the reference trajectory shows slight departures from the prescribed temperature limits, each below 0.5$\,^{\circ}\mathrm{C}$. In the IBDR case, the maximum hourly temperature difference is 1.04$\,^{\circ}\mathrm{C}$, and the reference trajectory remains within the prescribed temperature limits. The PBDR result therefore does not establish strict nonlinear feasibility without an additional margin. Two complementary responses are available when scheduling near such boundaries is required: the model can be tightened through a piecewise-linear power--current relation and partitioned McCormick envelopes, while execution can apply conservative margins and measurement-updated re-optimization, as discussed in the following Discussion section.}

\section{{\color{blue}Discussion}}
\label{sec:implementation}

{\color{blue}In a practical potline, the TMC-MILP would serve as a plant-level scheduling layer between the electricity-market or grid-dispatch interface and the existing smelter control system. Its output is the potline current trajectory, implemented through rectifier-current setpoints. It does not replace cell-level process control: local pot controllers would continue to regulate alumina feeding, ACD-related actions, and abnormal-event protection. Measurements and soft-sensor estimates from these lower control functions can provide the information needed to update the plant-level schedule, consistent with existing work on aluminum-reduction-process monitoring and control \cite{shi_multivariable_2020,shi_new_2021,shi_advanced_2023,wong_discretized_2021,mattioni_step_2024}.}

{\color{blue}The central practical question is how a nominal representative-cell schedule can be converted into a current command that remains safe for the individual cells in a potline. Cells may differ in their initial states and parameters, and some internal states can only be inferred from available measurements. In addition, the scheduling-level TMC model may differ from the actual cell because the lumped model does not resolve intra-cell spatial variations, while the MILP reformulation introduces a further approximation error. A safety margin therefore represents the distance between a predicted state and its safety limit after allowing for the possible deviation caused by these sources of error. During operation, the state estimate of each monitored cell would be updated as new measurements become available. If the predicted state of a cell approaches its limit, the relevant constraint should be tightened and the potline current trajectory re-optimized. Because all cells share the same potline current, the admissible trajectory must satisfy the tightened constraints of all monitored cells; the cell closest to a safety limit determines the available regulation range. Only the current setpoint for the next interval would be implemented, after which new measurements would be used to repeat the estimation and scheduling procedure.}

{\color{blue}Several sources of uncertainty can reduce the safety margin available to the scheduler, and they require different responses during operation. If uncertainty in the state estimate is dominant, the relevant measurements, soft sensors, or state observers should be improved. If parameter uncertainty is dominant, the parameters associated with the safety margin closest to its limit should receive priority attention. Thermal resistance, ambient conditions, and ledge-related quantities are most relevant when the temperature, superheat, or ledge-thickness margin is tight. Current efficiency and alumina-feed-related quantities are most relevant when the alumina-concentration margin is tight. When credible parameter ranges are available, the scheduling constraints can be required to remain satisfied across those parameter values through robust or scenario-based formulations \cite{langson_tube_2004,richards_robust_2006,bertsimas_theory_2011}. If the TMC predictions show a persistent systematic difference from plant observations, the scheduling-level model requires further calibration or refinement. If the audit in Section~\ref{sec:approx_error_audit} shows that MILP approximation error is the dominant source, the approximation itself should be improved by adding breakpoints, using a finer piecewise-linear power--current relation, or partitioning the McCormick envelopes, rather than simply enlarging the safety margin.}

\section{Conclusion}
\label{sec:conclusion}

This paper has developed a Thermal-Mass Coupled (TMC) dynamic model and a linearized MILP scheduling framework for aluminum smelting loads participating in demand response. {\color{blue}By retaining the thermal-mass mechanisms that determine the available load-modulation margin, the TMC model reduces the overly conservative flexibility estimate of ETP and the risk that the Linear model violates the thermal safety limits while remaining tractable for scheduling. In the illustrative PBDR and IBDR case studies, nonlinear reference-model assessment yields net economic improvements 122.4\% and 325.1\% higher, respectively, than ETP. The baseline-load and ToU-spread sensitivity analyses further show an economic advantage of TMC over ETP across the tested market and settlement settings. All MILP instances solve within 10\,s, supporting real-time re-optimization within the hourly scheduling interval.}

{\color{blue}Future work will extend the present framework in three directions: using industrial measurements to update cell states and calibrate model and parameters for uncertainty-aware multi-cell scheduling; coordinating the plant-level scheduler with lower-level cell controllers in closed-loop implementation; and broadening the market-side formulation beyond the current price-taking and nominal-baseline assumptions to account for price feedback, baseline-load uncertainty, and response-settlement risk.}

\section*{ACKNOWLEDGMENT}
Generative artificial intelligence was used to enhance grammar and readability.

\bibliographystyle{IEEEtran}
\bibliography{reference}
\end{document}